# Photochromism in single nitrogen-vacancy defect in diamond

T. Gaebel[1*], M. Domhan[1], C. Wittmann[1], I. Popa[1], F. Jelezko[1], J. Rabeau[2], A. Greentree[2], S. Prawer[2], E. Trajkov[3], P. R. Hemmer[3], and J. Wrachtrup[1]

[1] 3rd Physical Institute, University of Stuttgart, Stuttgart, Germany

[2] School of Physics, University of Melbourne, Victoria, Australia

[3] Department of Electrical Engineering, Texas A&M University, Texas, USA

* T. Gaebel

Email: t.gaebel@physik.uni-stuttgart.de

Fax: +49-711-685-5281

### Abstract

Photochromism in single nitrogen-vacancy optical centers in diamond is demonstrated. Time-resolved optical spectroscopy shows that intense irradiation at 514 nm switches the nitrogen-vacancy defects to the negative form. This defect state relaxes back to the neutral form under dark conditions. Temporal anticorrelation of photons emitted by the different charge states of the optical center unambiguously indicates that the nitrogen-vacancy defect accounts for both 575 nm and 638 nm emission bands. Possible mechanism of photochromism involving nitrogen donors is discussed.

### 1. Introduction

The photophysics of color centers in diamond has attracted interest during last decade because of their possible application in quantum information processing. Electron and nuclear spins associated with single nitrogen vacancy (N-V) defects in diamond are very promising qubits candidates because of the long coherence time and availability of reliable control via well established ESR-technique [1]. Recently, optical readout of single spins associated with N-V defects in diamond was demonstrated [2]. The coherent control of the electron spin of this defect and realization of two qubit conditional quantum gate was reported [3,4]. Additionally, the coupling to internal degrees of freedom, e.g., the hyperfine coupling to nuclei, like $^{13}$C and $^{14}$N, was investigated [5]. Although the N-V defect is one of the most studied color centers, it is remarkable that its basic photophysical properties (e.g., the charge state) are not well established yet. Such uncertainty is related to the fact that the charge state of defects in wide band gap materials is often affected by the proximity of donors (or acceptors). This explains why the concentration of isolated substitutional nitrogen (which is a major electron donor) defines the charge state of many optical centers in diamond. The best-known example is the vacancy, which exist in two stable charge states [6]. Photoinduced electron transfer can often change the charge state of a defect.

The N-V defect is a naturally occurring feature in diamond containing nitrogen. It consists of a substitutional nitrogen atom next to an adjacent vacancy in the diamond lattice. The defect can be produced by irradiation and subsequent annealing at temperatures above 550° C of nitrogen rich diamond. Due to the radiation damage, vacancies are created in the diamond lattice and the subsequent annealing treatment leads to migration of the vacancies towards nitrogen atoms creating N-V defects. Alternatively, the center can be produced in very pure diamond (with very low nitrogen concentration) by $N^+$ ion implantation [7,8]. Depending on other impurities in the close surrounding acting as electron acceptors or donors (e.g. nitrogen), the defect can occur in two charge states, neutral $(N-V)^0$ and negatively charged $(N-V)^-$.

The $(N-V)^-$ shows a Zero Phonon Line (ZPL) at 638 nm (1.945 eV) [9]. The properties of $(N-V)^-$ defect are well studied using hole burning, four-wave mixing and optically detected magnetic resonance techniques [10]. Based on these studies, a generally accepted six-electron model of the $(N-V)^-$ defect has been proposed. The energy level scheme consists of a ground state, which is a spin triplet of $^3A$ symmetry, an excited spin triplet state of $^3E$ symmetry, and an additional metastable singlet state ($^1A$). The ZPL absorption corresponds to $^3A \rightarrow ^3E$ transition. The main photophysical parameters, e.g., a large absorption cross section, the quantum yield close to unity ($\Phi = 0.99$), and a short excited-state lifetime ($\tau = 11.6$ ns) are suitable for single center detection using optical methods [1].

If electron donors are not available in the diamond lattice, the N-V center can be observed in its neutral charge state $(N-V)^0$. This charge state shows a ZPL lying at 575 nm (2.156 eV) and the effective electron spin of the neutral form of the defect is $S = ½$. The best indication, that both 575 nm and 638 nm optical bands belong to the different charge states of N-V center was given by heavy neutron irradiation of nitrogen-rich diamond [11]. A drastic change in the absorption spectra after heavy neutron irradiation and annealing was observed. The $(N-V)^-$ decreases and simultaneously sudden appearance of the $(N-V)^0$ were observed. This effect is explained by lowering of the Fermi level due to the irradiation and as a consequence the change of the charge state of the defect. In addition, the photochromic behavior of 575 nm and 638 nm emission bands was observed [12,13] and relation between them was studied [14,15]. In this paper we extend studies of photochromism to a *single* N-V defect. Observation of photochromic effect in a *single* defect provides unambiguous proof that both spectral lines belong to the different charge state of a single N-V defect.

## 2. Experiment

The sample used is a natural type IIa diamond (Element Six) implanted with 14 keV $N_2^+$ ions and subsequently annealed for 3 hours at 900° C. The ion dose used was approximately $2 \times 10^9$ $N_2^+$ ions $cm^{-2}$, which creates a N doped layer about 20nm thick in which the N concentration is approximately $6 \times 10^{14}$ $N/cm^3$ Type IIa diamond is very clean in respect to nitrogen impurities and most of the (N-V) defects created in this diamond originate from the implanted nitrogen. Experiments were performed using a home-built confocal microscope. The fluorescence light from the optical center was collected and directed to the detection channel, which includes a spectrometer and a Hanbury-Brown and Twiss interferometer. The excitation wavelength of 514 nm was blocked by a Notch 514 filter and a long-pass filter with cut-off at 570 nm. The interferometer was either used to measure the autocorrelation of the photons arriving at the two detectors, or, with additional filters (one detection arm a long-pass 667 nm for $(N-V)^-$ detection and the other detection arm a band-pass 585/80 nm for $(N-V)^0$ detection) to measure the cross-correlation between two charge states of defect. An acousto-optical modulator (AOM) was used for laser pulse shaping in time resolved experiments.

## 3. Results and Discussion

Single nitrogen vacancy defects have been detected via fluorescence using confocal optical microscope operating at room temperature. By choosing appropriate implantation dose the spacing between defects in the diamond was ensured to be larger than resolution of the optical microscope (0.3 µm). In this case the confocal image of the sample shows distinct fluorescence spots corresponding to fluorescence emission of the single N-V color centers. To check that N-V centers were observed, a fluorescence spectrum of a spot was taken (see Fig. 1). The spectrum shows two different ZPL, one at 575 nm for the $(N-V)^0$ and the second ZPL at 638 nm for the $(N-V)^-$ defect. There are also Stokes-shifted broad features corresponding to characteristic phonon frequencies of the center.

To check the number of fluorescence emitting centers in the laser focus, the autocorrelation function $g^{(2)}(\tau)$ has been measured following the procedure described previously [16]. The result of autocorrelation measurements performed in a spectral window corresponding to the $(N-V)^-$ state emission (see experimental section) are shown in Fig. 2a.

For the delay time $\tau = 0$ the function show a dip, indicating sub-Poissonian statistics of emitted light. This behavior is called antibunching and can be explained by the fact that a single quantum object can not emit two photons at the same time. Note that the contrast of antibunching dip scales as $1/n$ where $n$ is the number of defects in focus. Hence it gives an access to the number of centers studied. The correlation curve shown in Fig. 2a clearly shows the full contrast. Therefore only one single (N-V)$^-$ center is at the laser focus. The same result came out for the detection of the (N-V)$^0$ fluorescence (data not shown).

At first glance one might suggest from this result that in the fluorescence spot two defects: one (N-V)$^-$ defect and one (N-V)$^0$ defect are present. Such an interpretation would be consistent with both the fluorescence emission spectra (both zero phonon lines are visible) and the fluorescence correlation measurements. In this simple picture, both defects are independent one from another. As a consequence, even though the photons emitted by each defect show antibunching, no correlation would be expected between photons originating from different defects. On the other hand, if the light originates from a single NV center, then the NV$^-$ and NV$^0$ emission should be correlated.

Such cross-correlation measurements on photons, originating from two different charge states have been performed as a next step to test this conjecture. In this case, the fluorescence originating from the (N-V)$^0$ center was separated by appropriate optical filters and sent into one detection arm of the interferometer. The second arm was tuned to the spectral window corresponding to emission of (N-V)$^-$ defect. The data acquisition was started by detection of a photon in the (N-V)$^-$ arm and the time before next photon arrival in the (N-V)$^0$ arm was measured. Data presented in Fig. 2b is a histogram of such interphoton delays. The coincidence rate is close to zero for $\tau = 0$, indicating that the photons corresponding to different charge states of N-V center are anticorrelated. This leads to the following conclusions. First, the whole detected fluorescence (including (N-V)$^0$ and (N-V)$^-$ bands) stems from a single defect. Second, continuous switching between two charge states exists, because both bands are emitted by the same defect. Such photochromic behavior of N-V centers was previously reported for bulk samples where photoinduced changes occurred on a time scale of the order of seconds [12]. Although such photochromic behavior supports the assignment of both (575nm and 638 nm) bands to a single N-V center [12] later results has showed a lack of direct correlation between the two spectral bands and indicated the need for further verification [13]. The present observation of photochromic switching in a *single* defect resolves this uncertainty and establishes that both fluorescence bands (575 nm and 638 nm) originate from the same defect.

In order to further elucidate the nature of photochromic effect, time-resolved measurements have been performed. Pulsed laser excitation was employed to ascertain whether or not the change in the charge of the defect is photoinduced. First, the sample was allowed to relax in the dark state (i.e. the laser was switched off for 10 microseconds). Second, the laser pulse was applied and the fluorescence of both spectral bands (575 and 638 nm) was monitored independently. Fig. 3a shows the time-resolved fluorescence intensity for $(N-V)^0$ and $(N-V)^-$ bands. In the beginning of the laser pulse the fluorescence of $(N-V)^-$ state starts at a low level increasing exponentially towards a steady state intensity at longer illumination times. The situation for $(N-V)^0$ state is opposite, showing exponential decay of this state under illumination. This means that the switch from $(N-V)^0$ to $(N-V)^-$ state is photoinduced. Possible interpretation of this process is ionization of nitrogen donors present in the vicinity of N-V defect. The substitutional nitrogen donor level is 1.7 eV below the conduction band and it is ionized upon irradiation with 514 nm ( 2.41 eV) laser. Especially for molecular implantation (see experimental section), the probability to find the nitrogen donor in the close vicinity to N-V center is high.

The inverse process (conversion of $(N-V)^0$ state to $(N-V)^-$ state) occurs under dark conditions. The rate of this dark process has been determined by measuring the intensity of $(N-V)^0$ and $(N-V)^-$ states as a function of dark adaptation time (pulse sequence is shown in Fig. 3b). As expected, the behavior of two charge states shows opposite dynamics. Relaxation occurs towards $(N-V)^0$ state with a time constant of 2.1 µs. Significant spread of decay times (from 300 ns to 3.6 µs) has been observed for different defect centers. Such large variation of decay times may be related to the nature of the photochromic process. The photochromic behavior was observed only in natural type IIa diamonds, and not in synthetic type IIa crystals. The difference between synthetic and natural samples is not related to the position of NV centers energy levels within band gap, but rather to the charge states of nitrogen defects which can act as electron donors. It appears that the electron transfer rates between N-V centers and electron donor defect are distant dependent, and experimentally observed spread of relaxation rates for single N-V defects indicates the wide distribution of distances between N-V centers and nitrogen.

### 4. Conclusion

Optical spectroscopy of single nitrogen-vacancy defects ion diamonds unambiguously shows that 575 nm and 638 nm zero phonon lines belongs to the different charge states of nitrogen-vacancy defect in diamond. Upon 514 nm illumination the negative charge state appears to dominate. The defect can be converted into neutral charge state under dark conditions. Large spread of photochromic rates is consistent with the hypothesis of N donors being involved in the process. However, additional work on doped diamond crystals is required for precise identification of the electron donor responsible for the photochromic effect.

**Acknowledgement**

The work has been supported by the ARO, DFG via SFB/TR 21 and graduate college "Magnetische Resonanz", and the Landestiftung BW via the program "Atomoptik".

**Figure captions**

FIGURE 1: Fluorescence spectra of a spot of the confocal image. It shows two ZPLs, one at 575 nm indicating the (N-V)$^0$ defect and one at 638 nm for the (N-V)$^-$ defect in diamond. The characteristic pronounced phonon side wings of the centers are as well visible.

FIGURE 2: Fluorescence autocorrelation function of a single (N-V)$^-$ defect, measured with the Hanbury-Brown and Twiss interferometer filtering out the fluorescence from the (N-V)$^0$ defect. The measured coincidences were normalized and background corrected to obtain $g^{(2)}(\tau)$. The smooth line represent single exponential fit function (for details see reference [16])**(a)**. Coincidences between (N-V)$^-$ and (N-V)$^0$ center fluorescence. The distribution of interphoton delay times $\tau$ between a photon originating from the fluorescence of the (N-V)$^-$ center and a photon originating from the (N-V)$^0$ center was measured. The value for $\tau = 0$ indicates that both centers are one single quantum object **(b)**.

FIGURE 3: **(a)** The time dependence of the fluorescence intensity after switching on the excitation laser ($\lambda$ = 514 nm). **(b)** The fluorescence intensity of (N-V)$^0$ and (N-V)$^-$ states as a function of dark adaptation time $\tau$ for a single N-V center. The inset shows laser pulse sequence used.

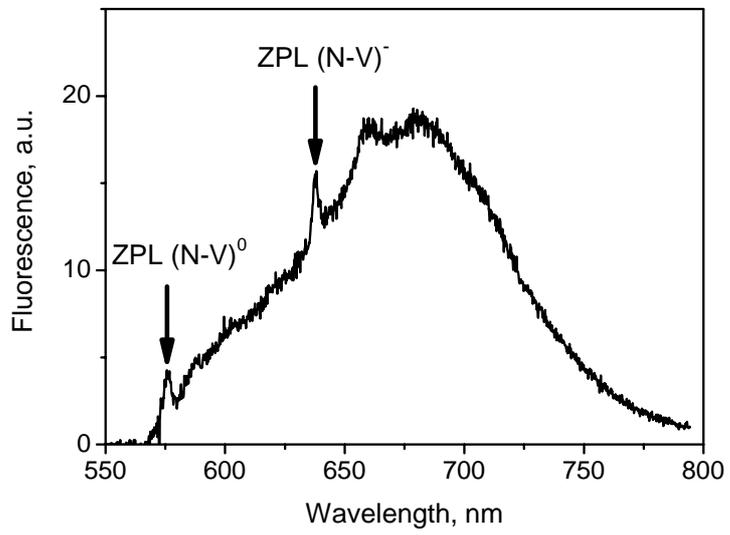

Figure 1

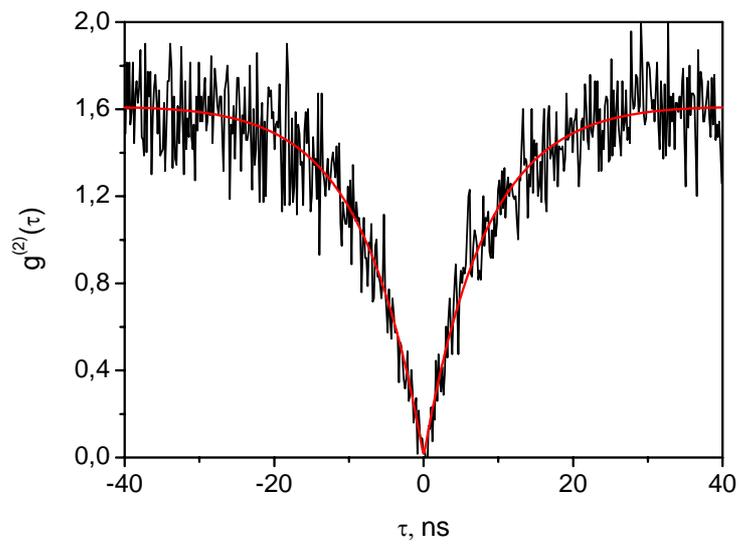

Figure 2 (a)

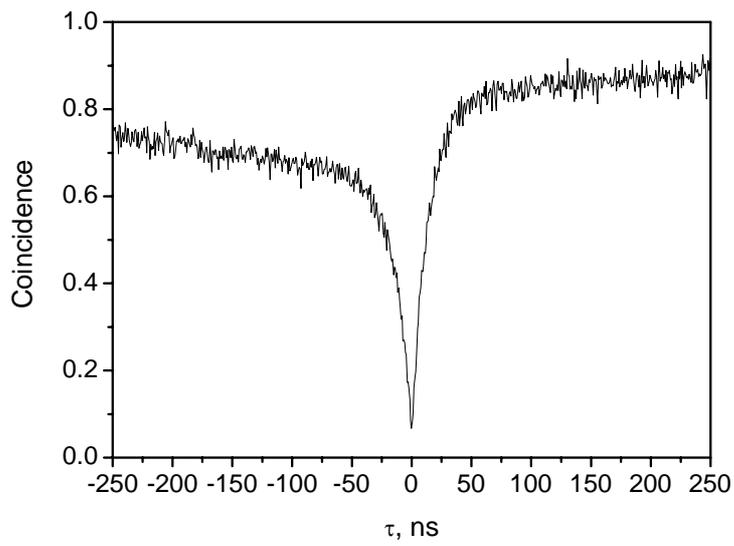

Figure 2 (b)

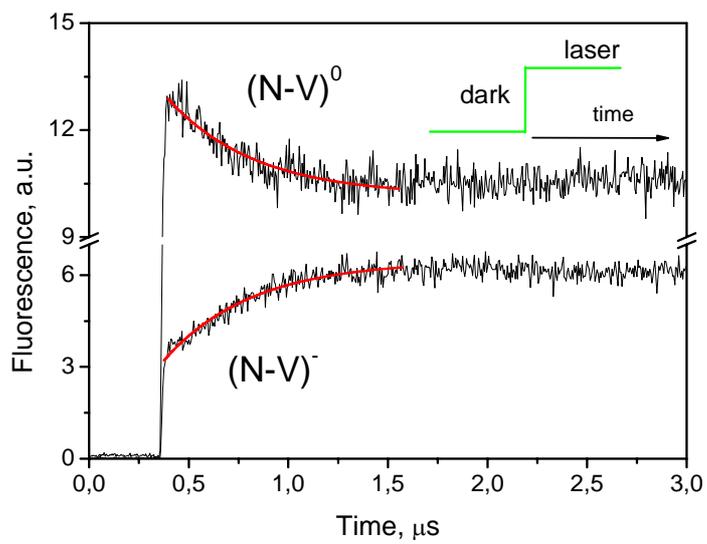

Figure 3 (a)

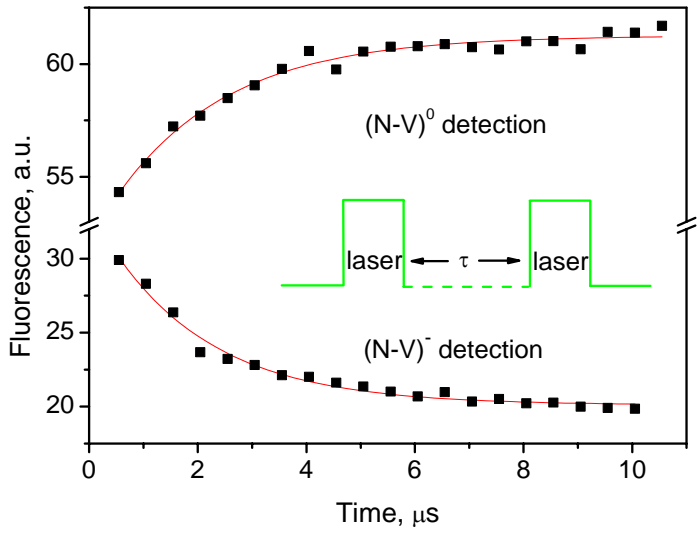

Figure 3 (b)